# 1320 nm Light Source From Deuterium Treated Silicon

SEREF KALEM  (Member, IEEE)

Electrical and Electronics Engineering Department, Bahcesehir University, Istanbul 34353, Turkey

This work was supported by the Scientific and Technological Research Council of Turkey (TUBITAK).

**ABSTRACT** We report an efficient room temperature photon source at 1320 nm telecommunication wavelength from nanostructured silicon surface. The activation of this light source was realized by treating the surface of Si wafer by vapor of heavy water ($D_2O$) containing a mixture of hydrofluoric and nitric acids. Treatment without deuterium generates an intense light emission band at the band-edge of Si, while the deuterium treatment alone creates a strong emission band at 1320 nm in the near infrared. It was found that the deuterium is actively involved in the formation of a nanostructured Si surface as evidenced from relative strength of the Si-O vibrational modes and presence of N-D bondings. The origin of this photon source was discussed in terms of oxygen related defect states and dislocations. The Si surface treated by Deuterium containing mixture exhibits a strong rectifying electrical activity as it is demonstrated by Schottky diodes fabricated on these wafers. Being compatible with mature silicon circuitry, the source may find applications in photonics and optoelectronics.

## I. INTRODUCTION

Silicon based optoelectronic devices and components are needed for monolithic integration of optical and electronic information processing units and interconnects. A silicon based near infrared (NIR) light source at transparent optical window is required for the fabrication of optoelectronic devices and on-chip communication. This type of NIR light source is expected to lead to a cost effective optoelectronic circuit production avoiding expensive heterogeneous integration of Si with optically active semiconductors such as III-V compound semiconductors and thus offering promising applications [1]–[5]. Photon emission from Si can only be activated through defects and quantum confinement effects [2], [3]. Due to its indirect bandgap nature, light emission from Si in NIR can usually be observed at low temperature, but without application possibility. The intensity of the source is relatively faint or not measurable at 300 K. One type of light emission source finds its origin at dislocations generated on the Si wafer [4], [5]. Devices fabricated from this type of source reveals rather broad bands having a number of emission peaks between 800–1800 nm [5], [6]. Si based NIR light emission can also be generated from dislocations, which are generated through misalignment during wafer bonding. But, light emission so obtained produces broad band light emission between 750 nm and 1900 nm [5]. NIR light source reported in this paper offers promising applications in the field of photonics and quantum technologies. The ability of the NIR light to pass through many semiconductor layer in Si circuitry without getting absorbed, new devices such as optical-electrical output transistors can be realized [7]. As an alternative method, we have shown with other research groups that the size reduction, for example in SiGe and Ge quantum dots, an NIR light can be observed even at 300 K [8], [9]. We learned also from the previous studies that an efficient band-edge emission can be obtained in strain induced wafers in addition to sub-gap emissions [10]. We know that semiconductor defects such as point defects and electronic impurities can be at the origin of the sub-band gap peaks in photoluminescence [11], [12]. Particularly, an optical emission at 1320 nm from sulfur doped crystalline Si was reported and identified as exciton photoluminescence from isoelectronic impurities [13]. This PL band was observed at low temperature and persisted up to T = 170 K.

Concerning the role of the deuterium, it is well known that deuterium has been used for the purposes of improving electronic quality of devices. For example, it was shown to improve dielectric properties of oxides in electronic devices [14]. The latter aimed at improving $HfO_2$ electronic quality by atomic layer deposition (ALD). It was also needed for the fabrication of low-loss components for optical communication. The electronic improvement of MOS structures have been annealed in deuterium atmosphere in order to improve their electronic quality [15]. It has been used together with HF vapor for surface cleaning and defect passivation on Si wafers and was found very effective in this process [16]. It is well known from earlier studies that deuterium is very effective in compensating dangling bonds, a typical source of electronic point defects, at the interfaces of Si and silicon dioxide. This effect has been resulted in the reduction of trap density improving I-V characteristics of complementary metal oxide semiconductor (CMOS) devices [17]. As a result of dangling bond compensation, it has been shown that deuterium incorporation into a semiconductor enhances radiative recombination by avoiding non-radiative degradation as evidenced from photoluminescence measurements [18]. Metal Oxide Semiconductor capacitors on Si wafers annealed in a deuterium atmosphere, revealed an effective enhancement in the performance of transistors [19]. Interesting features have been observed in $SiO_2$/Si so treated demonstrating a strong coupling of Si-D vibrations to Si-O-Si stretching modes [20].

Silicon as a semiconductor has excellent electronic properties, but the lack of useful light emission obliges semiconductor manufacturers to go with heterogeneous integration, what is a costly and difficult technology. However, light emission from Si is not straight forward process, requiring either down scaling of dimensions or defects induced by structural modifications. Previous studies on production of light emitting silicon wafers have rather been based on dislocation network generation through direct wafer bonding combined hydrogen induced layer splitting [21] revealing mid-infrared broad band emission. For the same reason, alternative forms of surface modification and phase changes of silicon have been under intense investigation. Among others, the discovery of a new tetragonal phase in Si [22], investigating fundamental mechanisms for brittle-to-ductile transition [23], [24]. Another example is the discovery of a new transition from Si-I to Si-VI via damaged Si [25] offering great potential for optoelectronic applications. Using a diamond wheel, complete amorphous surface can be fabricated on a single silicon crystal [27], which could not be manufactured by a conventional diamond wheel. These studies provide new insights and technologies for high performance devices used for optoelectronic industry. Our preliminary surface deformation studies have shown infrared emission at room temperature from silicon, which was treated with heavy water containing vapors of a chemical mixture (HF : $HNO_3$ : $D_2O$) [27]. The present work is related to a detailed systematic study of novel light source, which is emitting an efficient light in the mid-infrared ranging from 900 nm to 1700 nm having full band width at half maximum FWHM of around 180 nm with a quantum efficiency of greater than 50% at room temperature. These findings can lead to exciting opportunities for an indirect bandgap Si having a chance to introduce related optical and electronic components to its advanced microelectronics manufacturing line.

As-grown or as-treated wafers following treatment of wafer surface by vapors of a chemical mixture have been investigated by physical, structural and electrical measurements. Photoluminescence (PL), spectroscopic ellipsometry (SE), Fourier-transformed infrared (FTIR) spectroscopy, scanning electron microscopy (SEM), energy dispersive spectroscopy (EDS) and current-voltage (I-V) characteristics. From EDS analysis, we have found that there is 24 at.% of oxygen in the surface layer. This amount of oxygen concentration is considered to be relatively high when we compare it with a wafer treated by hydrogen (about 5 at.%) instead of deuterium. Metal contact for electrical measurements used Au on Ti/$SiO_x$ deposited on surface and the device so-produced on this structure reveals a Schottky diode characteristics as evidenced from a strong rectifying diode behavior.

The process of surface treatment on Si consists of exposing wafers to heavy water $D_2O$ containing HF : $HNO_3$ acid vapor, which results in the formation of ammonium silicon hexafluoride $(ND_4)_2SiF_6$ on silicon surface [27]. This way of treating wafer surface leads to a light emitting structure with characteristic single and high quantum efficiency emission peak at 1320 nm. This work differs from previous studies with a single emission band at room temperature (RT) in the mid-infrared region as well as with its production method using a heavy water. It is a new method of sensitizing Si wafer in order to induce NIR light emission using relatively simple and low cost technology. Having these properties, our method is promising for fabricating light emitting devices such and sensors.

## II. EXPERIMENTAL RESULTS AND DISCUSSION
### A. DEUTERIUM TREATMENT
To produce a light emitting surface, silicon wafer is treated by deuterium oxide $D_2O$ containing mixture of HF:$HNO_3$:$D_2O$, which produces acid vapor, where the concentration ratio of HF in the solution can be around 40%–50% and concentration of $HNO_3$ solution should be 60%–70%. The processing of wafer has been carried out at room temperature in process chamber made of teflon. The concentration of hydrofluoric acid (HF) and nitric acid) $HNO_3$ used in our experiments were about 48% and 65% by weight (mass percent), respectively. The chemical quality of the solutions is of semiconductor grade. HF:$HNO_3$:$D_2O$ chemical solution consists of concentrations of 2–7 unit volume from HF, 1–12 unit volume from $HNO_3$ and 1–6 unit volume of 5–20 Ohm-cm and 5–10 Ohm-cm for p-type and n-type, respectively. Si wafers with <100> and Si <111> crystal orientations have been used in the surface modification process.

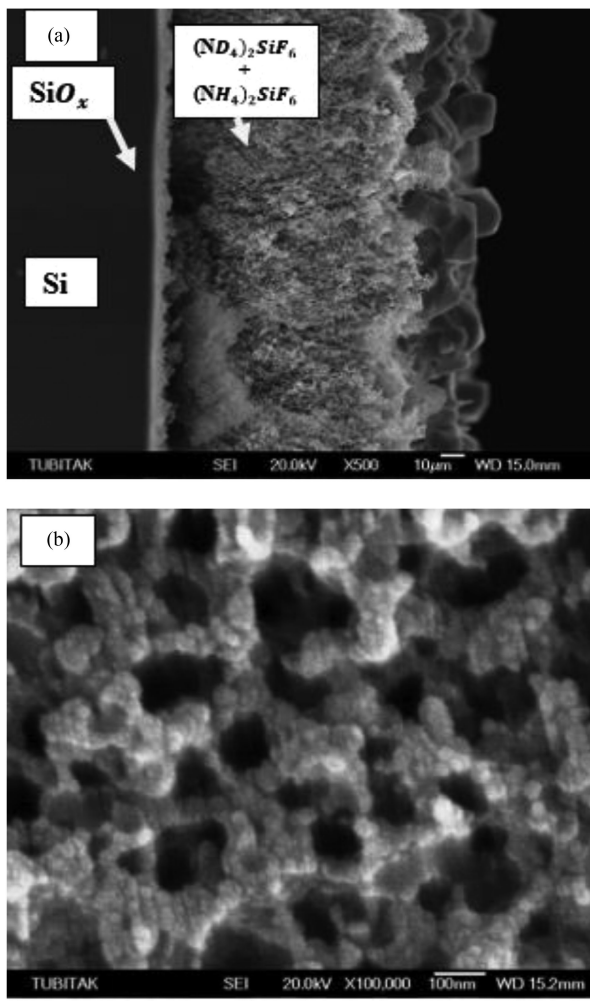

**FIG. 1.** (A) SEM micrograph taken from the cross-section of the layer. (B) Surface morphology following deuterium treatment. EDS analysis shows that surface has around 24 at.% oxygen and 76 at.% Si. The depth of the holes (dark areas) or cavities can be up to 500 nm as evidenced from the cross-sectional SEM images.

### A. MICROSTRUCTURE

The cross-sectional and surface structure micrographs as taken by SEM from the wafer, which was treated by vapor of $HF:HNO_3:D_2O$ acid mixture are shown in Fig. 1(A) and (B), respectively. The experimental setup used for producing deuterium treated wafers have been reported elsewhere [28]. As a result of the treatment, a dielectric surface structure is formed, consisting of deuterium ammonium silicon hexafluoride dielectric layer $(ND_4)_2SiF_6$ (hereafter DASH) with sub-oxide interface as shown in Fig. 1(A). The dielectric thin film contains also $(NH_4)_2SiF_6$ (hereafter ASH) as evidenced from the FTIR analysis as shown in Fig. 2. The DASH layer can be removed through rinsing of the wafer in deionized (DI) water. After removal of this layer a surface structure or the interface layer becomes visible as shown in Fig. 1(B). As evidenced from the SEM surface micrographs, the surface consists of nano-sized grains of Si nanocrystals, which are encapsulated by silicon sub-oxide $SiO_x$. SEM micrographs shows the presence of nanometer sized grains, which are bound together as shown in Fig. 1(B). This type of ordering also forms large cavities (dark areas) with diameters ranging from 10 nm to 100 nm. The depth of these surface structures ranges from about 300 nm to 500 nm as evidenced by cross-sectional SEM images. The production of nm sized structures on silicon wafer has been realized using a number of methods, which generates a light emission in the visible spectral range, mainly a broad band at around 600 nm. Despite an extensive study of nanostructured silicon, the peak emission at 1320 nm using our method offers novel features [29]–[31]. Therefore, deuterium treatment of wafers generates interesting features in silicon wafer surface, thus offering very rich physical and electrical properties, that is worth to study.

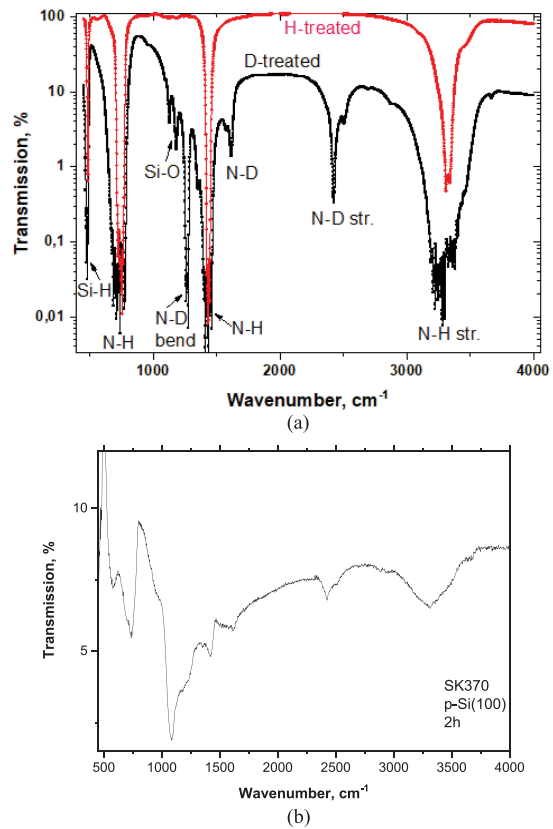

**FIG. 2.** (A) Evidence of deuterium incorporation. Absorption bands of N-D and N-H bonds. Si-O stretching modes are located at about 1100 cm-1. Frequency of N-D vibrations is shifted by 1.376 corresponding to mass ratio of D and H atoms. The figure compares the spectra of H and D treated wafers. (B) Early stages of deuterium treated p-Si, where wafer was exposed to D2O containing mixture for 2 hours. The strength of Si-O-Si stretching modes at 1085 cm-1 is the strongest among other modes.

### B. DEVICE FABRICATION

After the deuterium treatment process has been done, wafers were rinsed in DI water and which was followed by dipping in 5% HF solution for 10 seconds to etch oxide layer from the surface of wafer. For metallization, a 5 nm thick Ti layer was deposited on wafer, which was followed by an evaporation of 20 nm Au electrod through shadow mask. For the bottom electrode, an aluminum electrod of 100 nm was sputtered on the back side of the wafer.

## C. PHYSICAL AND ELECTRICAL CHARACTERIZATION
### 1) VIBRATIONAL MODES ANALYSIS

Localized vibrational modes as investigated by FTIR analysis provided crucial insight into the microstructure of as-grown wafers. These investigations reveal that deuterium interacts with Si and nitrogen atoms forming bonding within the DASH matrix, as evidenced by the observation of related bonding in FTIR spectra. A typical Fourier Transformed Infra Red (FTIR) spectra of the DASH wafers of p-Si (111), which were processed for 6 hours (HF : $HNO_3$ : $D_2O$) are shown in Fig. 2. The same figure shows also the spectrum of H-treated (HF : $HNO_3$ : $H_2O$) wafer for comparison. FTIR investigation performed on as-grown samples revealed some modifications in microstructure of wafer, which has been treated by $D_2O$ containing solution. We find that stretching mode (Si-O-Si) at 1090 $cm^{-1}$ is relatively intense at the beginning of the treatment, indicating that oxidation process is strong at the early stages of treatment of Si with deuterium containing vapor. Relative weakness of Si-O-Si vibrations at later stages of the process (6 hours of treatment) as shown in Fig. 2(A) supports the presence of strong oxidation process at early stages of the treatment. As an example, the spectrum for the early stages of deuterium treated p-Si is shown in Fig. 2(B), where the wafer was exposed to $D_2O$ containing mixture for 2 hours. We observe that the strength of the Si-O-Si stretching modes at 1085 $cm^{-1}$ is the strongest among other modes. From these studies, it is clear that atomic masses, frequency shifting of 1.376 with respect to N-H bands at 3132 $cm^{-1}$ correspond to these new vibrations. Hence these vibrations were attributed to deuterium related modes. This ratio of 1.376 is actually equal to the ratio of masses of N-D and N-H bonds. Si-O stretching modes are located at about 1100 $cm^{-1}$. Frequency of N-D vibrations is shifted by 1.376 corresponding to mass ratio of D and H atoms. The figure compares the spectra of H and D treated wafers. deuterium and hydrogen atoms, that is $m_D/m_H = 1.376$. Replacement of H atoms by D atoms is expected to shift the N-H mode frequency down to 2425 $cm^{-1}$. These bands at 2425 $cm^{-1}$ and 1260 $cm^{-1}$ can be attributed to stretching and bending bands of N-D vibrations, respectively. On the other hand, FTIR spectra also exhibit Si-O stretching modes at 1092 $cm^{-1}$, 1128 $cm^{-1}$ and 1183 $cm^{-1}$. These bands can be attributed to oxide vibrations as originating from Si-O-Si stretching modes from previous studies [47]. Moreover, we observe a doublet of strong intensity at 1263 $cm^{-1}$ and 1272 $cm^{-1}$, and identify these bands as the coupling of closely spaced N-D stretching mode pairs [32]. The band at lower frequencies at 485 $cm^{-1}$ is present in both deuterium and hydrogen treated samples, therefore it is identified as originating from N-H wagging vibrational modes. From these studies, it is clear that deuterium has a significant role in the formation of novel surface structure as shown in Fig. 1. These studies are a strong proof of the presence of deuterium atoms, which are well attached to the ASH matrix. The demonstration of these bonds is evidenced by observing N-D related bonding modes. These modes, which have been observed in Fig. 2 at 2524 $cm^{-1}$ and 2425 $cm^{-1}$ were attributed to N-D and N-$D_2$ stretching bands. The presence of additional bands as compared to ASH samples can be readily observed as they are intense absorptions. When we consider the ratio of atomic masses, frequency shifting of 1.376 with respect to N-H bands at 3132 $cm^{-1}$ correspond to these new vibrations. Hence these vibrations were attributed to deuterium related modes. This ratio of 1.376 is actually equal to the ratio of masses of deuterium and hydrogen atoms, that is $m_D/m_H = 1.376$. The replacement of H atoms by D atoms is expected to shift the N-H mode frequency down to 2425 $cm^{-1}$. These bands at 2425 $cm^{-1}$ and 1260 $cm^{-1}$ can be attributed to stretching and bending bands of N-D vibrations, respectively. On the other hand, FTIR spectra also exhibit Si-O stretching modes at 1092 $cm^{-1}$, 1128 $cm^{-1}$ and 1183 $cm^{-1}$. These bands can be attributed to oxide vibrations as originating from Si-O-Si stretching modes from previous studies [47]. Moreover, we observe a doublet of strong intensity at 1263 $cm^{-1}$ and 1272 $cm^{-1}$, and identify these bands as the coupling of closely spaced N-D stretching mode pairs [32]. The band at lower frequencies at 485 $cm^{-1}$ is present in both deuterium and hydrogen treated samples, therefore it is identified as originating from N-H wagging vibrational modes. There is a very thin (few 100 nm thick depending on the process conditions) sub-oxide (SiOx) layer on Si treated by $D_2O$ containing acid vapor, where **x** was estimated to be close to 2.0 from the peak position of Si-O vibration [32]. This oxide layer can be up to few hundred nm thick. SEM-EDS analysis show that oxygen concentration within this layer would be up to 25%. We assume the interface layer is an oxide (SiOx) with embedded Si quantum dots or nano size crystals. This amount of oxygen is relatively large when we compare it with that of hydrogen that is about 5 at.%. This result shows that nano crystalline grains on surface are encapsulated by silicon oxide layer, thus probably favoring also quantum size effects in optical processes.

## D. SPECTROSCOPIC ELLIPSOMETRY

The presence of Si nanometer size surface structures and their effects on optical properties are supported by the data obtained from SE analysis. Fig. 3 shows the second derivative of dielectric function $\varepsilon(\omega)$. Despite the surface roughness, the main critical point corresponding to energy band structure of Si can still be observed. The critical points (CP) can be identified as energy points belonging to Si and nanometer sized structures, wherein quantum confinement effects are expected. Energy point at 3.43 eV is likely due to E1 critical point in the bulk Si. E1 CP, direct bandgap edge, is broad and sharpest, indicating importance of direct band optical transitions. The transition energy at 4.65 eV is somewhat higher than the bulk E2 CP, which could be attributed to quantum confinement related transition. We observe that around 600–700 nm interference effects are quenched. This region can be indicative of the presence of silicon oxide related transition. The PL peak at about 1.8 eV can be due to such type of transition and related radiative recombination at this optical region. From the photoluminescence peak energies, we estimate that an effective nano crystal size of 1 nm to 5 nm following earlier work. From these findings we assume that the surface layer consists of Si quantum dots embedded in oxide matrix. But, the radiative recombination pathways of these PL bands could involve defects at around Si nanocrystals. Decay times of about 5 ns measured in these wafers are in support of the assumptions [3].

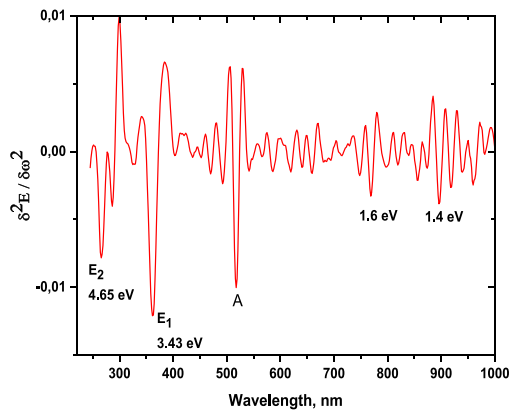

**FIG. 3.** Spectroscopic ellipsometry indicates critical energy points of the surface layer formed during the deuterium treatment. Letter A indicates an experimental artifact.

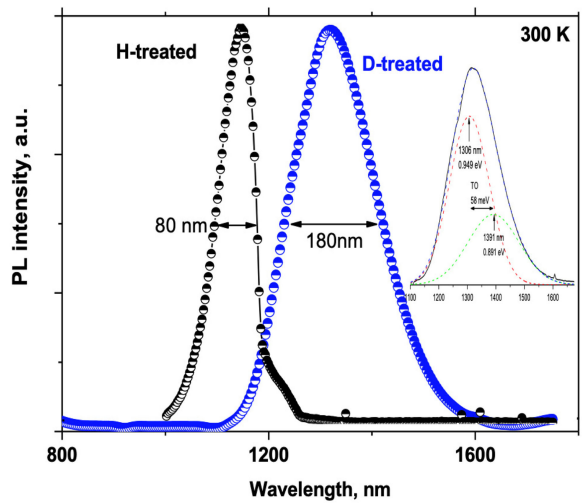

**FIG. 4.** Room temperature photoluminescence at 1320 nm from Si treated by vapor of HF:HNO$_3$:D$_2$O acid. Light emission was compared with H-treated wafer having also a strong intensity photon emission band at 1145 nm.

### III. 1320 NM LIGHT SOURCE

Our investigations indicate that the treatment of wafers by D$_2$O containing vapor leads to exciting tunable optical properties. Fig. 4 is typical photoluminescence emission spectrum obtained from a wafer, which was treated with heavy water as compared with a wafer treated by vapor containing hydrogen. The excitation of as-grown sample used an Argon ion laser line of 488 nm at 300 K. When wafer is treated only by hydrogen containing vapor, the PL spectra exhibit a very strong intensity single peak at 1145 nm, which was attributed to the band edge emission of silicon. With the treatment using Deuterium, the band edge emission disappears and a very strong single peak at 1320 nm appears in the spectra. We find that the emission band at 1320 nm has an FWHM of around 180 nm, which is more than twice that of the band edge emission (80 nm). This type of emission could be due to D3 type dislocations as previously observed at misaligned wafer bonding process [34] and dislocation loops [35] [41] or from impurities [11]. Nevertheless, the PL is very broad and measured at low temperature, which is accompanied by satellite emissions at the band edge and D1 dislocation emission at 1500 nm. Nevertheless, the identification of the origin of the band at 1320 nm is not straightforward. It is located around D3 dislocation luminescence line of Si but D3 was identified as the phonon replica of D4 line thus appearing together at low temperature [4]. Actually, we have shown that this band can be de-convoluted to 1306 nm (0.949 eV) and 1391 nm (0.891 eV) peaks with a band distance of 58 meV corresponding to transverse optical TO-phonon energy as shown at the insert of Fig. 4. The component at 1306 can be attributed to the PL originating from no-phonon recombination of electrons and holes in 1D band of the D4 dislocation line [36]–[38]. Then the 1320 nm band can be attributed to the TO phonon assisted replicas of the dislocation line D4 related luminescence. But, the new energetic positions of the lines are not anymore in line with the reported values of dislocation lines. Therefore, we rule out the possibility of a phonon replica. Moreover, a recent work has demonstrated an inconsistent luminescence intensity relation between D3 and D4 lines suggesting that different origins can be considered for D3 dislocation line [39].

The electronic band structure of Si/SiOx interface can be schematically represented as shown in Fig. 5 for a p-type Si wafer. Oxide encapsulation creates an hole accumulation region at the interface, where the holes are recombined by either electrons at interface traps or by those at the 1D band of a dislocation line in Si. At the front layer, e-h pairs which are generated in the quantum wells are recombined radiatively leading to visible light. There is also an effective recombination pathways through the interface states between Si and SiOx. The probability of drift or tunneling through the thin oxide layer into the bulk silicon can be significant. But also the contribution from the defect states at the surface or dislocations or stacking faults near surface or at Si nanowires can be effective in NIR light emission at around 1350 nm. An infrared photoluminescence was observed in Si nanowires with high density of stacking faults [44]. But emission is rather very broad and ranges from 900 nm up to 1650 nm. Other example shows that hydrogen ion implantation could lead to optically active defects, which can generate 1300 nm light emission in Si [45]. Moreover, it was reported that hydrogenation enhances particularly D3 band at around 1300 nm in Si [46]. The nature of these defects however is not yet clear, but

**TABLE 1.** Localized Vibrational Modes Frequencies

| Sample | TO Si-Si N-Hwag. [45] | N-H/Si-N str. [45] | Si-O str. Asym. Str.[46] | N-D str. [37] | N-H bend. /deform.[45] | N-D bend. [45] | N-D bend. [45] | N-H sym. str. [45] |
|---|---|---|---|---|---|---|---|---|
| H6hrs | 482vs | 744vs | 1104vw/1184vw |  | 1437vs |  |  | 3329vs |
| D6hrs | 469s<br>485s | 714vs<br>774vs | 1090vw<br>1128m/1183m | 1265vs | 1432vs | 1613m | 2425s | 3275vs |
| D2hrs | 456m, 475sh | 734vs | 1085vs , 1200sh |  | 1421m | 1615w | 2424s | 3315vs |

Localized vibrational bands observed in deuterium treated samples as compared to samples treated without deuterium. Also shown are the bands of a wafer treated with deuterium for 2 hours, where H6hrs and D6hrs represent samples treated with hydrogen and deuterium for 6 hours, respectively. Relative intensities are indicated as vs(very strong), s(strong), m(medium), w(weak), vw(very weak), sh(shoulder).

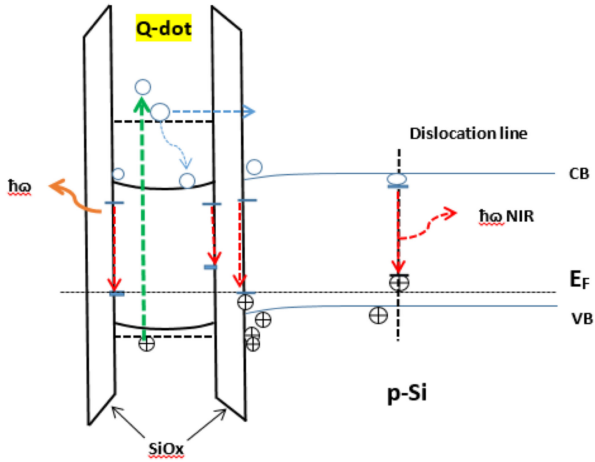

**FIG. 5.** Schematic energy band structure of an oxide encapsulated silicon at the interface of a p-type Si wafer. Visible radiative emission is due to the recombination of excited carriers through defect states at SiOx/Si interface. Some of the excited carriers can diffuse or tunnel through the oxide contributing to the NIR emission in Si. Dislocation line indicates where e-h pairs are recombined leading to efficient 1320 nm emission at defect states in stacking faults. Q-dot indicates quantum dot.

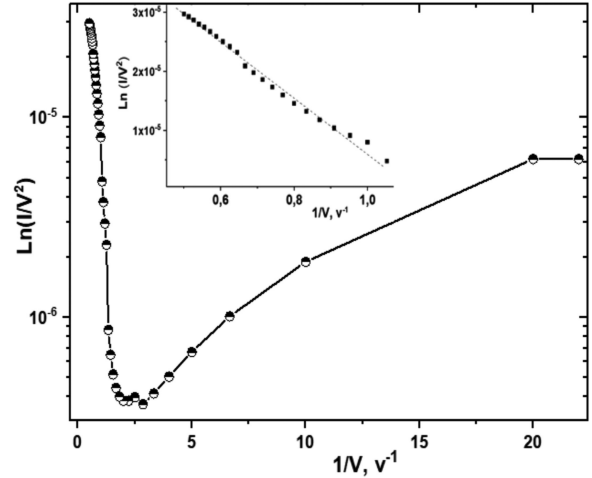

**FIG. 6.** Current-voltage plot of an Au/Si device probed by Au electrod on p-Si (111) 16 h 15 min. The current for all Schottky diodes exhibit rectifying behavior having a threshold voltage of about 4 V. The insert shows the linear relationship between Ln (1/V²) and 1/V at low bias indicating Fowler-Nordheim tunneling.

the compensation of dangling bonds in both examples is likely the main reason for the photoluminescence enhancement.

### A. DEVICE PROPERTIES

Fig. 6 shows a typical electrical characteristic of the heterojunction in terms of Ln $(I/V^2)$ versus 1/V indicating Fowler-Nordheim type of conduction. The p-type Si wafer was treated by Deuterium for 16 hours and 15 minutes. A 5 nm thick Ti layer was deposited, which followed by evaporation of a 20 nm Au layer through direct deposition shadow mask to form top electrode. For the bottom electrode, an aluminum layer of 100 nm was deposited on back side of the wafer using sputtering tool. We have performed electrical characterization measurements to determine the device characteristics using current-voltage behavior. We find that deuterium treated samples exhibit rectifying characteristics of a Schottky diode. From these analysis, we observe that Au/Deuterium treated Si/p-Si device revealed a threshold of less than 0.4 V at forward bias with a breakdown voltage greater than −10 V at reverse bias. We observe series resistance of 83 Ω and 101 Ω at 1 V and 0.5 V, respectively. These low values can be attributed to the nature of surface structure consisting of some tapered ends easing carrier injection. The ideality factor can be estimated using this expression $n = \frac{q}{kT}(\frac{\partial V}{\partial lnI})$ leading to 3.9, a relatively high value relative to bulk Si. This high **n** value can be related to disorder induced transport at interface between the Au electrode and Deuterium treated surface layer. The figure indicates that there exists two different electrical conduction mechanisms, resulting from direct tunneling and Fowler-Nordheim (F-N) tunneling when V > 2.2 V, 1/V = 0.45 V$^{-1}$. The linear relationship between Ln $(I/V^2)$ and 1/V is indicative of the F-N type of tunneling as shown at the insert in Fig. 6. From slope of the plot, barrier height was estimated to be about 2.08 eV.

### IV. CONCLUSION

In summary, we have shown that with deuterium treatment Si wafer can be turned into a very effective light source generating at the near infrared telecommunication wavelength. The modification of Si wafer surface using the vapor of heavy water $D_2O$ containing HF:HNO$_3$ transforms Si surface into a dielectric layer and an SiOx interface wherein Si nanosized crystals embedded. The Si wafer surface so modified reveals

VOLUME 1, 2020 93

rich optical characteristics and a very efficient light emission at 1320 nm. This band has some overlap with D3 dislocation PL line suggesting a D3 related origin for the peak. However, oxidation induced stacking faults or a deep level at around 200 meV above the valance band could also be at the origin of the defect states leading to the photon source. Deuterium treatment induces an heavy oxidation leading to likely SiOx precipitates, which nucleate the stacking fault of oxygen interstitials. As an alternative explanation, charges that are trapped at interface states between oxide encapsulation and silicon can lower potential barrier in nano-structured Si/SiOx device leading to observed emission band. Diodes produced on deuterium treated wafers exhibit very strong rectifying characteristics at relatively low thresholds and high breakdown voltages, thus offering the possibility of operation devices at high temperatures. Detailed studies are required to determine the role of the disorder generated surface/interface states and dislocations in generating light at 1320 nm. The nature of these states and dislocations would of particular interest in clarifying the origin of the NIR photon source. This novelty can lead to a broad range of applications where Si based monolithic optoelectronics and photonics circuitries are in great demand in information processing and telecommunication technologies.

## REFERENCES

[1] M. Hochberg and T. Baehr-Jones, "Toward fabless silicon photonics," *Nature Photon.*, vol. 4, p. 1, 2010.
[2] L. Pavesi, "Silicon-based light sources for silicon integrated circuits," *Adv. Opt. Technologies*, ID416926, p. 1, 2008.
[3] T. Schmidt *et al.*, *Physical Rev. B*, vol. 86, 2012, Art. no. 125302.
[4] C. Krause *et al.*, "Properties of strong luminescence at 0.93 eV in solar grade silicon," *Solid State Phenomena*, vol. 83, pp. 205–206, 2014.
[5] M. Kittler, M. Reiche, T. Argirov, W. Seifert, and X. Yu," *Physica Status Solidi A*, vol. 203, no. 802, 2006.
[6] R. P. Schmid *et al.*, *Phys. Status Solidi A*, vol. 208, p. 888, 2011.
[7] S Saito, D. Hisamoto, and H. Shimizu," *Appl. Phys. Lett.*, vol. 89, 2007, Art. no. 163504.
[8] S. Kalem *et al.*, "Near-IR photoluminescence from Si/Ge nanowire grown silicon wafers," *Appl. Phys. A*, vol. 112, p. 561, 2013.
[9] G. Abstreiter *et al.*, *Semicond. Sci. Technol.*, vol. 11, 1996, Art. no. 1521.
[10] S. Kalem *et al.*, *Nanotechnology*, vol. 22, p. 1, 2011.
[11] O. King and D. G. Hall, *Physical Rev. B*, vol. 50, 1994, Art. no. 10661.
[12] V.G. Talalaev *et al.*, *Nanoscale Res. Lett.*, vol. 1, p. 137, 2006.
[13] T.G. Brown and D.G. Hall," *Appl. Phys. Lett.*, vol. 49, p. 245, 1986.
[14] H.-H. Tseng, *IEDM Tech. Dig.* vol. 83, 2003.
[15] T. Kundu," *IEEE Trans. Device Mater. Rel.*, vol. 6, p. 288, 2006.
[16] J. Mizsei, A. E. Pap, K. Gillemot, and G. Battistig, *Appl. Surface Sci.*, vol. 256, 2010, Art. no. 5765.
[17] D. Misra and RK Jarwal, "Interface hardening with deuterium implantation," *J. Electrochem. Soc.*, vol. 149, 2002, Paper G446.
[18] T Matsumoto, A I Belogorokhov, L I Belogorokhova, Y. Masumoto, and E. A Zhukov," *Nanotechnology*, vol. 11, p. 340, 2000.
[19] J. W. Lyding, K. Hess, and I. C. Kizilyalli," *Appl. Phys. Lett.*, *vol*. 68, 1996, Art. no. 2526.
[20] Z. Chen, J. Guo, and P. Ong," *Appl. Phys. Lett.*, vol. 83, 2003, Art. no. 2151.
[21] M. Kittler and M. Reiche," *Adv. Eng Mat*, vol. 11, p. 249, 2009.
[22] B. Wang *et al.*, "New deformation-induced nanostructure in silicon," *Nano Lett.*, vol. 18, no. 7, pp. 4611–4617, 2018
[23] Z. Zhang, B. Wang, R. Kang, B. Zhang, and D. Guo, "Changes in surface layer of silicon wafers from diamond scratching," *CIRP Annals-Manuf. Technol.*, vol. 64, pp. 349–352, 2015.
[24] Z. Zhang, D. Guo, B. Wang, R. Kang, and B. Zhang, "A novel approach of high speed scratching on silicon wafers at nanoscale depths of cut," *Scientific Rep.*, | 5:16395 | DOI: 10.1038/srep16395.
[25] Z. Zhang *et al.*, "Deformation induced new pathways in silicon," *Nanoscale*, vol. 11, pp. 9862–9868, 2019.
[26] Z. Zhang *et al.*, "Deformation induced complete amorphization, at nanoscale in a bulk silicon," *AIP Adv.*, vol. 9, 2019, Art. no. 025101.
[27] S. Kalem, "Controlling photon emission from silicon for photonic applications," *Proc. of SPIE* vol. 9364, 2015, Art. no. 93641P.
[28] S. Kalem and O Yavuz," *Opt. Express*, vol. 6, no. 7, 2000.
[29] S. Kalem *et al.*, *Nanotechnology*, vol. 20, 2009, Art. no. 445303.
[30] E. Fritsch *et al.*, *Appl. Phys.*, vol. 90, 2001, , Art. no. 4777.
[31] M. Cheung *et al.*, *SPIE J. Nanophotonics*, vol. 5, 2011, Art. no. 053503.
[32] I. A. Oxton *et al.*, "Infrared spectra of the ammonium ion in crystals," *J. Physical Chem.*, vol. 80, 1976, Art. no. 1212.
[33] N. Tomozeiu, *Optoelectronics - Materials and Techniques*, Prof. P. Predeep (Ed.), ISBN: 978-953-307-276-0, InTech, 2011..
[34] D. Mankovics, R. P. Schmid, T. V. Arguirov, and M. Kittler, "Dislocation related photoluminescence imaging of mc-Si wafers at room temperature," *Crystal Res. Technol.*, vol. 47, no. 11, pp. 1148–1152, 2012.
[35] T. Hoang *et al.*, *IEEE Trans. Electron Devices*, vol. 54, 2007, Art. no. 1860
[36] V. Kveder and M. Kittler, "Dislocations in silicon and D-band luminescence for infrared light emitters," *Mater. Sci. Forum*, vol. 29, p. 590, 2008.
[37] M. Kittler *et al.*, *Mater. Sci. Eng. C*, vol. 27, 2007, Art. no. 1252.
[38] T. Mchelidze *et al.*, *Solid State Phenomena*, vol. 156-158, p. 567, 2009.
[39] H. T. Nguyen *et al.*, *IEEE J. Photovolt.*, vol. 5, p. 799, 2015.
[40] T. V. Arguirov, PhD Thesis, BTU Cottbus, 2007.
[41] C. Krause, "Investigation of particular crystal defects in solar silicon materials using electron beam techniques," Thesis, Cottbus, 2015.
[42] M. Y. Chou, M. L. Cohen, and S. G. Louie, *Physical Rev. B*, voll. 32, 1985, Art. no. 7979.
[43] L. F. Mattheiss and J. R. Patel, *Physical Rev. B*, vol. 23, 1981), Art. no. 5384.
[44] Y. Li *et al.*, *J. Nanoscale*, vol. 7, 2015, Art. no. 1601.
[45] S. Boninelli *et al.*, "Hydrogen induced optically active defects in silicon photonic nanocavities," *Opt. Express*, vol. 22, 2014, Art. no. 8843.
[46] S. Kalem, "Defect studies in strain relaxed Si1-xGex alloys," *Turk J. Phys.*, vol. 37, p. 275, 2013.
[47] S. Kalem, "Synthesis of ammonium silicon fluoride cryptocrystals on Silicon," *Appl. Surface Sci.*, vol. 236, pp. 336–341, 2004.
[48] N. Nagai and H. Hashimoto, FTIR-ATR study of depth profile of SiO2 ultra-thin films," *Appl. Surface Sci.*, vol. 172, p. 307, 2001.